%% file: tsrml_2022.tex
\documentclass{article}

\usepackage[final]{tsrml_2022}

\usepackage[utf8]{inputenc} % allow utf-8 input
\usepackage[T1]{fontenc}    % use 8-bit T1 fonts
\usepackage{hyperref}       % hyperlinks
\usepackage{url}            % simple URL typesetting
\usepackage{booktabs}       % professional-quality tables
\usepackage{amsfonts}       % blackboard math symbols
\usepackage{nicefrac}       % compact symbols for 1/2, etc.
\usepackage{microtype} % microtypography
\usepackage{subfigure}
\usepackage{xcolor}         % colors
\input{math_commands.tex}

\usepackage{wrapfig,lipsum,booktabs}

\usepackage{multirow,mathtools } 
\title{Formatting Instructions For TSRML 2022}

\author{%
  Jinghan Jia \\
  Michigan State University \\
  \texttt{jiajingh@msu.edu} \\
  \And 
  Mingyi Hong \\
  University of Minnesota \\
  \texttt{mhong@umn.edu} \\
  \And
  Yimeng Zhang \\
  Michigan State University \\ 
  \texttt{zhan1853@msu.edu} \\
\And
  Mehmet Akçakaya \\
  University of Minnesota \\ 
  \texttt{akcakaya@umn.edu} \\
  \And
  Sijia Liu \\
  Michigan State University \\
  \texttt{liusiji5@msu.edu}
}

\begin{document}

\title{On the Robustness of deep learning-based MRI Reconstruction to image transformations}
\maketitle

\begin{abstract}
Although deep learning (DL) has received much attention in   accelerated magnetic resonance imaging (MRI), recent studies show that tiny input perturbations may lead to instabilities of DL-based MRI reconstruction models. However, the approaches of robustifying these models are underdeveloped. Compared to image classification,   it could be much more challenging to achieve a robust MRI image reconstruction network considering its    regression-based learning objective,    limited amount of training data, and   lack of efficient robustness metrics.  To circumvent the above limitations, our work  revisits the problem of  DL-based image reconstruction through the lens of robust machine learning. We  find  a new instability source of MRI image reconstruction, \textit{i.e.}, the lack of reconstruction robustness against spatial transformations of an input,
 \textit{e.g.}, rotation and cutout. 
Inspired by this new robustness metric, we develop a   robustness-aware image reconstruction method that can defend against both pixel-wise adversarial perturbations as well as spatial transformations. Extensive experiments are also conducted to demonstrate the effectiveness of our proposed approaches. 
\end{abstract}

\section{Introduction}
\label{sec:intro}
Medical image reconstruction is a known problem in magnetic resonance imaging (MRI). The key challenge lies in solving an inverse problem so as to  enable the reconstruction of a 
high-quality medical image from sub-sampled measurements that  are readily be obtained from medical devices \citep{lustig2008compressed}.
% acquisition time.
The conventional  approach for solving the MRI image reconstruction problem commonly resorted to sparse signal processing techniques \citep{lustig2008compressed,yang2010fast,huang2011efficient}, \textit{e.g.}, compressed sensing. The key enabling technique is to leverage  sparsity-inducing optimization \citep{bach2011optimization}, \textit{e.g.}, the least-squared problem penalized by $\ell_1$ norm. 
In contrast to the direct optimization-based approaches,  an increasing amount of recent works proposed the deep learning (DL)-based MRI image reconstruction methods \citep{zhu2018image,liang2020deep,sriram2020end}. With aid of DL models, the problem of image reconstruction is then solved as supervised learning problem 
in which the reconstruction error between the `predicted' high-quality image and the train-time `reference' image is minimized.  In this work, our research falls into the category of DL-based MRI image reconstruction.

The DL-based image reconstruction approaches    can be roughly divided into three types. First, U-Net \citep{ronneberger2015u} and its many variants \citep{han2018framing,lee2018deep} follow the Encoder-Decoder based network architecture.   
Second, learning-to-optimize (L2O), which leverages a recurrent neural network (RNN) to mimic an iterative optimization algorithm in problem solving \citep{chen2017learning}, was adopted to unroll the optimization flow of sparse signal recovery for image reconstruction \citep{yang2016deep,aggarwal2018modl,hosseini2020dense}
Third, in addition to supervised learning, state-of-the-art unsupervised learning approaches  were leveraged to tackle medical image reconstruction in the scarce data regime \citep{hu2021self,fabian2021data}.

In spite of the rapid growth in DL for medical image reconstruction, 
nearly all existing works focused on tackling the problem of enhancing image \textit{reconstruction accuracy}. However, a series of recent works \citep{antun2020instabilities,zhang2021instabilities} showed that the DL-based image reconstruction network suffers a \textit{weakness}: the lack of adversarial robustness. Specifically, 
it has been shown in \cite{antun2020instabilities} that 
an imagery input, at the presence of  adversarial perturbations (which  are  designed via adversarial learning, known as adversarial attack generation), can significantly hamper the  image reconstruction accuracy even if these perturbations are unnoticeable to human eyes. We refer readers to Fig.\,2(b) for a reconstructed image example using an DL-based approach when facing adversarial input perturbations. 
Prior to the finding of the lack of adversarial robustness in \textit{image reconstruction}, the problem of adversarial learning has been extensively studied in the domain of \textit{image classification}
 \citep{goodfellow2014explaining,carlini2017towards,croce2020reliable}. 
 To defend such adversarial perturbations, the min-max optimization-based adversarial training (AT) \citep{madry2017towards} provides a principled defensive method to robustify image classifiers. By contrast, few work studied the problem of improving
   adversarial robustness for deep image reconstruction networks, except \cite{raj2020improving, caliva2020adversarial} which were mainly built upon AT.  Different from the existing work, our \textbf{contributions} are summarized below.

\noindent $\bullet$ 
We show that in addition to adversarial perturbation, deep MRI image reconstruction models are also vulnerable to  tiny {input spatial transformations}.

\noindent $\bullet$ 
We develop a generalized adversarial training  (GAT)  approach that promotes robustness against both {adversarial input perturbations} and {input spatial transformations}.

\noindent $\bullet$ 
We demonstrate the effectiveness of GAT on the  MRI dataset \citep{aggarwal2018modl}, showing our improvement over the conventional standard training and adversarial training.

\section{Preliminaries}
\label{sec:PRELIMINARIES}
\begin{figure}[htbp]
\centering
\subfigure[\footnotesize{ input $\mathbf z$}]{
\begin{minipage}[t]{0.33\linewidth}
\centering
\includegraphics[width=1.0in]{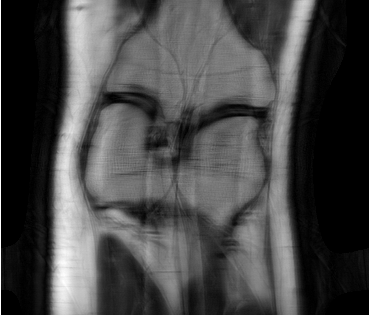}
\end{minipage}%
}%   
\subfigure[\footnotesize{output $f_{\boldsymbol \theta_{\mathrm{NT}}}(\mathbf z)$ }]{
\begin{minipage}[t]{0.30\linewidth}
\centering
\includegraphics[width=1.0in]{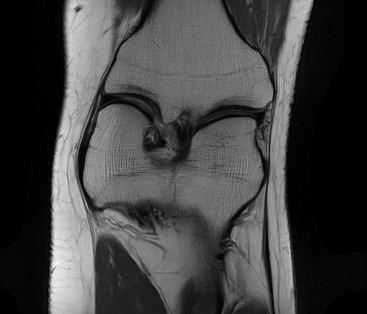}
\end{minipage}%
}%   
\subfigure[\footnotesize{ground truth $x^\ast$}]{
\begin{minipage}[t]{0.30\linewidth}
\centering
\includegraphics[width=1.0in]{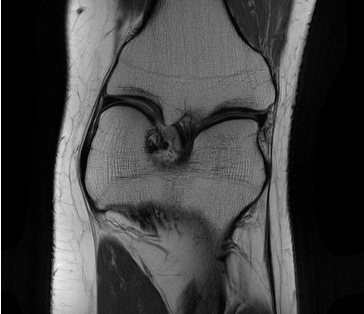}
\end{minipage}%
}%

\centering
\vspace*{-3mm}

\caption{\footnotesize{Example of multi-coil MRI image reconstruction using  Model-based deep learning  architecture \citep{aggarwal2018modl}  under fastMRI knee \citep{knoll2020fastmri} dataset by solving \eqref{eq: nm_train}. We refer readers to {Sec.\,\ref{sec:experiments}} for more implementation details.}}
\label{fig: recon_visual_normal}
\end{figure}

In this section, we  provide a brief background on the problem of accelerated MRI acquisition, and the deep learning (DL) based MRI reconstruction method.
\vspace*{-4mm}
\paragraph*{Accelerated multi-coil MRI acquisition}
\label{ssec:Accelerated multi-coil MRI acquisition}
In MRI , measurement of a patient's anatomy is acquired in the  Fourier-domain, also known as the $k$-space, through receiver coils. In the setup of multi-coil MRI acquisition, each coil  will capture a different region of a targeted image, and will be assigned with  a complex-valued sensitive map, noted by $\mathbf S_i$. As a result, 
  the $k$-space measurement $\mathbf y \in \sC ^n$ of a complex-valued ground truth image $\mathbf x^\ast \in \sC ^n$ is given by :
\begin{align}
 \label{eq: multi-MRI}
    \begin{array}{l}
\displaystyle \mathbf y_i = \mathbf{P}_\Omega\mathcal F \mathbf S_i  \mathbf{x}^\ast + \mathbf  n_i,\quad  i=1,...,N,  
    \end{array}
\end{align}
where $\mathbf N$ is the number of coils,
$\mathbf P_\Omega$ is a binary sampling mask that will be introduced later,
$\mathcal F$ is the two-dimensional Fourier-transform, and $\mathbf n_i \in \sC^n$ is the additive noise arising in the measurement process. 
In \eqref{eq: multi-MRI}, the use of  $\mathbf P_\Omega$ is for the purpose of \textit{accelerating} the multi-coil MRI acquisition, since
obtaining the full-sampled measurements in the $k$-space is quite time-consuming. The $0$s in $\mathbf P_\Omega$ represents the frequencies that will not be sampled.
We refer readers to \cite{deshmane2012parallel} for more details on  multi-coil MRI acquisition.
Eventually, the  multi-coil MRI acquisition process can be formulated as the forward mapping below \citep{pruessmann2001advances}
\begin{align}
 \label{eq: under-sampled}
    \begin{array}{l}
\displaystyle{ \mathbf y_\Omega} = \mathbf{A}_\Omega \mathbf x^\ast+\mathbf n % = \mathbf A \circ \mathbf x^* + \mathbf z,  
    \end{array}
\end{align}
where $ \mathbf y_\Omega$ denotes the sub-sampled multi-coil $k$-space measurements,
$\mathcal A_\Omega$ denotes the forward encoding operator which concatenates $\mathbf{P}_\Omega\mathcal F \mathbf S_i$ across all coils.
\vspace*{-4mm}
\paragraph*{MRI reconstruction problem}
\label{ssec:subhead}
The goal of MRI reconstruction is to estimate the original image $\mathbf x^*$ from the  $k$-space measurement  $ \mathbf y_\Omega$. The conventional solution to   dealing with  the MRI reconstruction problem is to treat the problem as
 an  inverse problem and  then calls the regularized least-squared method \citep{pruessmann2001advances,sutton2003fast} .
 %\SL{[refs]}
However, the state-of-the-art reconstruction performance is achieved by a learning-based approach, especially for using deep models. 
In this paradigm, we aim to learn a   reconstruction network $f_{\boldsymbol \theta}$, parameterized by $\boldsymbol \theta$,  under a training data set $\mathcal D$.
In the training set, the data `feature' is given by the sub-sampled image $\mathbf z$ acquired from the $k$-space measurement directly, namely, $\mathbf z = \mathbf A_{\Omega}^H \mathbf y_\Omega$, and the data `label' is given by the original image $\mathbf x^*$. The rationale behind using $\mathbf z$ rather than $\mathbf y_\Omega$ is that the mapping from $\mathbf z$ and $\mathbf x^*$ can be readily achieved using a denoising-based neural network \citep{aggarwal2018modl} as the input resolution and the output resolution remain the same. 
%\SL{[ref]}
The deep learning (DL) based MRI reconstruction then seeks the optimal   network  parameters $\boldsymbol \theta$ by minimizing a certain training loss $\ell$ over the training dataset $\mathcal D$:
% {\small
% \vspace*{-5mm}
% {
\begin{align}
 \label{eq: nm_train}
    \begin{array}{l}
\displaystyle \min_{\boldsymbol{\theta}} ~\mathbb E_{(\mathbf z,  \mathbf x^\ast) \in \mathcal D } \left [     ~  \ell( f_{\boldsymbol \theta}(\mathbf z), \mathbf x^\ast) \right ],
    \end{array}
\end{align}%}}
where  recall that $f_{\boldsymbol \theta}(\mathbf z)$ and $\mathbf x^\ast$ correspond to the network output image and the reference fully-sampled image, respectively.
In practice,  a
normalized $\ell_1$-$\ell_2$ loss \citep{knoll2020deep} is commonly used to specify $\ell$, which is given by
% \SL{[xxx]}
\begin{align}\label{eq: loss_l12}
   \ell( f_{\boldsymbol \theta}(\mathbf z), \mathbf x^\ast) = \frac{
   \| f_{\boldsymbol \theta}(\mathbf z) -  \mathbf x^\ast \|_2}{ \|  \mathbf x^\ast \|_2} + \frac{
   \| f_{\boldsymbol \theta}(\mathbf z) -  \mathbf x^\ast \|_1}{ \|  \mathbf x^\ast \|_1}.
 \end{align}
Fig.\,\ref{fig: recon_visual_normal} shows 
% an example to demonstrate 
the input sub-sampled image $\mathbf z$, the  reconstructed image  obtained from the normally trained  model ${\boldsymbol \theta_{\mathrm{NT}}}$ by solving problem \eqref{eq: nm_train}, and the full-sampled reference image  $\mathbf x^*$. 

\vspace*{-3mm}
\section{The Lack of Robustness for DL-Based MRI Reconstruction}
\begin{figure}[htbp]
\centering
\subfigure[\footnotesize{ $f_{\boldsymbol \theta_{\mathrm{NT}}}(\mathbf z)$}]{
\begin{minipage}[t]{0.30\linewidth}
\centering
\includegraphics[width=1.0in]{Figures/recon_ATAX.png}
\end{minipage}%
}
\subfigure[\footnotesize{$f_{\boldsymbol \theta_{\mathrm{NT}}}(\mathbf z+ \boldsymbol \delta)$}]{
\begin{minipage}[t]{0.30\linewidth}
\centering
\includegraphics[width=1.0in]{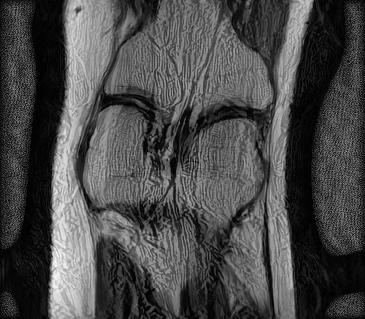}
\end{minipage}
}
\subfigure[\footnotesize{ground truth $\mathbf x^\ast$}]{
\begin{minipage}[t]{0.30\linewidth}
\centering
\includegraphics[width=1.0in]{Figures/ref.png}
\end{minipage}%
}% 
\vspace*{-1mm}
\centering
\caption{\footnotesize{Example of MRI image reconstruction based on the benign input $\mathbf z$ and the adversarial input $\mathbf z + \boldsymbol \delta$, versus the reference image $\mathbf{\mathbf x^*}$.}}
\vspace*{-4mm}
\label{fig: recon_adv_input}
\end{figure}

In this section, we will demonstrate two vulnerabilities of the DL-based MRI   reconstruction network to input `perturbations'. First, we echo the finding of \cite{antun2020instabilities} that 
DL-based MRI reconstruction is oversensitive to \textit{adversarial input perturbations}, which are unnoticeable but carefully crafted to worsen the reconstruction accuracy. Second, beyond \cite{antun2020instabilities}, we find that DL-based MRI reconstruction also lacks robustness against \textit{input spatial transformations}, \textit{e.g.}, flipping, cutting-out, and rotation, even to a small degree. To the best of our knowledge, the vulnerability of DL-based MRI reconstruction to spatial transformations of inputs is found for the first time. 
\paragraph*{Norm-constrained adversarial input perturbations}
Given a well-trained reconstruction network $f_{\boldsymbol \theta_{\mathrm{NT}}}$ from \eqref{eq: nm_train}, the  norm-constrained adversarial attack was first established by \cite{antun2020instabilities}  in the context of MRI image reconstruction. Following the same spirit of adversarial attack in image classification \citep{goodfellow2014explaining}, the adversarially perturbed input is formulated as $\mathbf z + \boldsymbol \delta$, where $\boldsymbol \delta$ denotes adversarial perturbations. The adversary then optimizes $\boldsymbol \delta$ to degrade the reconstruction performance. To this end, one can solve the   optimization problem \citep{antun2020instabilities},
\begin{align}\label{eq: atk_adv}
    \displaystyle \max_{\| \boldsymbol \delta \|_\infty \leq \epsilon} \ell(f_{\boldsymbol \theta_{\mathrm{NT}}}(\mathbf z + \boldsymbol \delta), f_{\boldsymbol \theta_{\mathrm{NT}}}(\mathbf z)),
\end{align}
where $\|  \cdot \|_\infty $ denotes the $\ell_\infty $ norm, $\epsilon > 0$ is the perturbation budget, and $\ell$ signifies the reconstruction loss, \textit{e.g.}, given by \eqref{eq: loss_l12}. The rational behind   \eqref{eq: atk_adv} is to learn $\boldsymbol \delta$ so as to enlarge the discrepancy between the model outputs at the perturbed input and the original input, respectively.
To solve the optimization problem \eqref{eq: atk_adv},  the standard projected gradient descent (PGD) \citep{madry2017towards} is commonly used. In Fig.\,\ref{fig: recon_adv_input}, we compare  the image reconstruction results (when facing the benign input and the adversarial perturbed input, respectively) with the full-sampled reference image $\mathbf x^*$. To generate the adversarial perturbations $\boldsymbol \delta$, we use PGD to solve \eqref{eq: atk_adv} with $\epsilon = 0.03/255$. 
\vspace*{2mm}

\begin{figure}[htbp]
\centering
\subfigure[\footnotesize{input $\mathbf z^\prime$}]{
\begin{minipage}[t]{0.30\linewidth}
\centering
\includegraphics[width=1.0in]{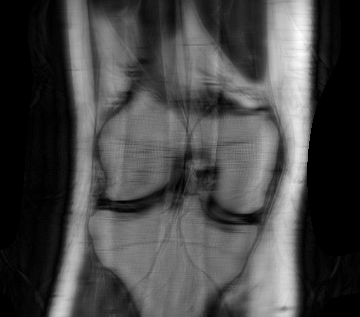}
\end{minipage}%
}%   
\subfigure[\footnotesize{$f_{\boldsymbol \theta_{\mathrm{NT}}}(T(\mathbf z))$ }]{
\begin{minipage}[t]{0.30\linewidth}
\centering
\includegraphics[width=1.0in]{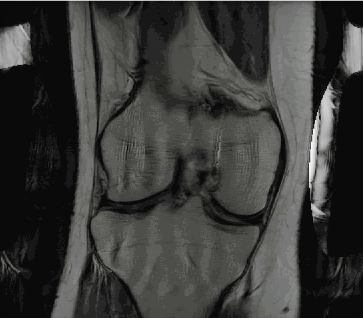}
\end{minipage}%
}%   
\subfigure[\footnotesize{ground truth $T(\mathbf x^\ast)$}]{
\begin{minipage}[t]{0.30\linewidth}
\centering
\includegraphics[width=1.0in]{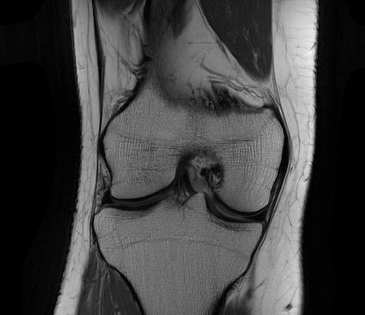}
\end{minipage}%
}% 
\centering
\vspace*{-4mm}
\caption{\footnotesize{Example of MRI reconstruction at the spatially transformed input (by rotating $180 ^{\circ}$), the output image, and the reference image.}}
\vspace*{-4mm}

\label{fig: recon_tranformation}
\end{figure}

\paragraph*{Spatial transformation-based input perturbations.}
We next introduce a   non-norm-constrained adversarial perturbations by leveraging input spatial transformations, \textit{e.g.}, flipping, cutting-out, and rotation. {We ask}: \textit{Will DL-based MRI reconstruction network be robust against heuristics-based input transformations?}
Different from adversarial perturbations that are determined by solving \eqref{eq: atk_adv}, the input transformations are randomly generated.
Let $ T: \sC^n \to \sC^n $ denote a transformation operation,
the perturbed input is then given by
%the transformation attack will generate attack examples as following:
\begin{align}
 \label{eq: input transformation}
    \begin{array}{l}
\displaystyle { \mathbf z'} = T(\mathbf z).
    \end{array}
\end{align}
Associated with \eqref{eq: input transformation}, the desired reference image will become $  T(\mathbf x^*)$. 
Given the normally trained network $f_{\boldsymbol \theta_{\mathrm{NT}}}$, we find that there exists a large discrepancy between $f_{\boldsymbol \theta_{\mathrm{NT}}} (T(\mathbf z))$ and its reference image $  T(\mathbf x^*)$. An example is shown in Fig.\,\ref{fig: recon_tranformation}, where a rotation of $180 ^{\circ}$ is applied to the input image $\mathbf z$. Moreover, by comparing $f_{\boldsymbol \theta_{\mathrm{NT}}}(T(\mathbf z))$ with $f_{\boldsymbol \theta_{\mathrm{NT}}}(\mathbf z + \boldsymbol \delta)$ in Fig.\,\ref{fig: recon_adv_input}, we   observe that
DL-based MRI reconstruction lacks robustness against not only optimization-based adversarial perturbations but also heuristics-based input transformations. 

\section{Towards Unified Robustness of MRI Reconstruction}
\label{sec:pagestyle}
\begin{figure}[htbp]
\centering
\subfigure[\footnotesize{ $f_{\boldsymbol \theta_{\mathrm{NT}}}(\mathbf z + \boldsymbol \delta)$}]{
\begin{minipage}[t]{0.30\linewidth}
\centering
\includegraphics[width=1.0in]{Figures/recon_adv_ATAX.png}
\end{minipage}%
}%   
\subfigure[\footnotesize{$f_{\boldsymbol \theta_{\mathrm{AT}}}(\mathbf z + \boldsymbol \delta)$}]{
\begin{minipage}[t]{0.30\linewidth}
\centering
\includegraphics[width=1.0in]{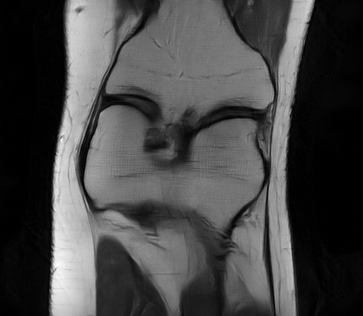}
\end{minipage}
}
\subfigure[\footnotesize{$ f_{\boldsymbol \theta_{\mathrm{AT}}}(T(\mathbf z))$ }]{
\begin{minipage}[t]{0.30\linewidth}
\centering
\includegraphics[width=1.0in]{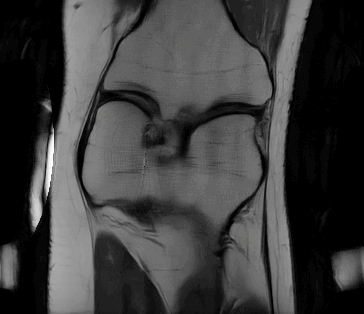}
\end{minipage}%
}%   
\vspace*{-3mm}
\caption{\footnotesize{Comparison of MRI reconstruction: (a) The normally trained network $f_{\boldsymbol \theta_{\mathrm{NT}}}$ at the benign input, (b) the AT model $f_{\boldsymbol \theta_{\mathrm{AT}}}$ at the adversarially perturbed input, (c) the AT model $f_{\boldsymbol \theta_{\mathrm{AT}}}$ at the transformed   input (by rotation of $180^{\circ}$). we rotate it $180^{\circ}$ for easier comparison with others.
} }
\vspace*{-5mm}
\label{fig: recon_visual_AT}
\end{figure}

In this section, we will develop novel training protocols to acquire robust DL-based MRI reconstruction networks against both adversarial perturbations and input transformations.
To defend adversarial perturbations, the min-max optimization (MMO)-based adversarial training (AT) recipe will be used following the same spirit of robust image classification \citep{madry2017towards}. However, we will show that AT lacks efficacy to improve robustness against input transformations.  Spurred by that, we will then propose a generalized AT approach that can incorporate data augmentations for `free', namely, without needing additional data annotations. 
\paragraph*{Adversarial training (AT)}
\label{ssec:Ad}
AT adopts MMO as the  algorithmic backbone, where the \textit{worst-case (maximum)} training loss is \textit{minimized} by incorporating a {synthesized attack} during model training. For robust image classification, AT has been poised {the only}  effective defense method  \citep{athalye2018obfuscated} against the adversarial input perturbations. Based on MMO, problem \eqref{eq: nm_train} can be modified as 
\begin{align}
 \label{eq: nm_train_MMO}
    \begin{array}{l}
\displaystyle \min_{\boldsymbol{\theta}} ~\mathbb E_{(\mathbf z,  \mathbf x_\ast) \in \mathcal D } \max_{\|\boldsymbol \delta\|_\infty \leq \epsilon } \left [     ~  \ell( f_{\boldsymbol \theta}(\mathbf z + \boldsymbol \delta), \mathbf x^\ast) \right ],
    \end{array}
\end{align}%}}
where the training loss is given by \eqref{eq: loss_l12}. Compared to  \eqref{eq: nm_train}, the attack generation problem \eqref{eq: atk_adv} is infused into \eqref{eq: nm_train_MMO} as an inner maximization problem. To solve problem \eqref{eq: nm_train_MMO}, the alternating gradient ascent-descent  method \citep{madry2017towards} is commonly used. 
For ease of presentation, let $\boldsymbol \theta_{\mathrm{AT}}$ denote the model parameters obtained from \eqref{eq: nm_train_MMO}. This is in contrast to $\boldsymbol \theta_{\mathrm{NT}}$ obtained by the normal training over \eqref{eq: nm_train}.
In Fig.\,\ref{fig: recon_visual_AT}, we demonstrate an example to compare the reconstruction performance of $f_{\boldsymbol \theta_{\mathrm{AT}}} (\mathbf z + \boldsymbol \delta )$ and that of $f_{\boldsymbol \theta_{\mathrm{NT}}} (\mathbf z + \boldsymbol \delta)$ when facing adversarial perturbations. Meanwhile, 
we show the performance of these different models against input transformations. As we can see, AT largely enhances the robustness of MRI reconstruction against adversarial input perturbations.  However, 
it is insufficient to robustify the reconstruction network against the input transformations; see Fig.\,\ref{fig: recon_visual_AT}-(c).

\paragraph*{Generalized AT (GAT) with data augmentation}
\label{ssec:Ad_aug}
Spurred by the lack of robustness of AT against input transformations, we ask if the transformed data can be infused into the AT framework. 
Meanwhile, such integration of AT with data augmentation is expected to be as simple as possible without needing additional data annotations. To this end, we can augment the training data set $\mathcal D$ with the newly paired transformed data $\tilde{\mathcal D} = \{ T(\mathbf z), T(\mathbf x^*)  \}_{(\mathbf z, \mathbf x^*) \in \mathcal D}$. In this context, we will generalize AT  into the following regularized optimization framework 
\begin{align}
      \hspace*{-5mm}
  \begin{array}{l}
\displaystyle \min_{\boldsymbol{\theta}} ~\underbrace{ \mathbb E_{(\mathbf z,  \mathbf x_\ast) \in \mathcal D \cup \tilde{\mathcal D}}  \left [     \ell( f_{\boldsymbol \theta}(\mathbf z), \mathbf x^\ast) \right ] }_\text{augmented normal training}
+ \underbrace{ \gamma \displaystyle  ~\mathbb E_{(\mathbf z,  \mathbf x_\ast) \in \mathcal D \cup \tilde{\mathcal D}} \max_{\|\boldsymbol \delta\|_\infty \leq \epsilon } \left [     ~  \ell( f_{\boldsymbol \theta}(\mathbf z + \boldsymbol \delta), f_{\boldsymbol \theta}(\mathbf z)) \right ]}_\text{robustness regularization}, 
    \end{array}
    \hspace*{-5mm}
 \label{eq: nm_train_MMO_aug}
\end{align}%}}
where $\gamma > 0$ is a regularization parameter to strike a balance between normal training and robust training. 
Different from \eqref{eq: nm_train_MMO}, we explicitly introduce an augmented standard training objective to improve the reconstruction accuracy when facing input transformations, and we introduce a `label-free' robust regularization that penalizes the discrepancy between the normal reconstruction and the adversarial reconstruction.
This robustness regularization is commonly used to tradeoff the standard model accuracy and the model robustness \citep{zhang2019theoretically}.

\begin{figure}[htbp]
\vspace*{-4mm}
\centering
\subfigure[\footnotesize{ $f_{\boldsymbol \theta_{\mathrm{GAT}}}(\mathbf z)$}]{
\begin{minipage}[t]{0.30\linewidth}
\centering
\includegraphics[width=1.0in]{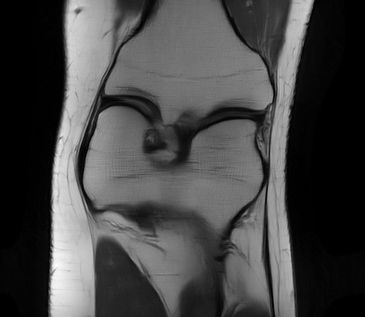}
\end{minipage}%
}%   
\subfigure[\footnotesize{$f_{\boldsymbol \theta_{\mathrm{GAT}}}(T(\mathbf z))$ }]{
\begin{minipage}[t]{0.30\linewidth}
\centering
\includegraphics[width=1.0in]{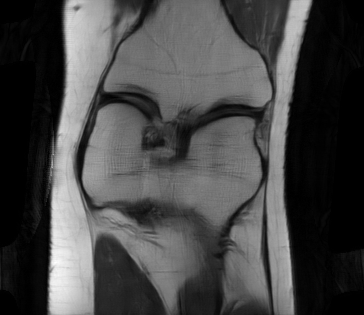}
\end{minipage}%
}%   
\subfigure[\footnotesize{$f_{\boldsymbol \theta_{\mathrm{GAT}}}(\mathbf z+ \boldsymbol \delta)$}]{
\begin{minipage}[t]{0.30\linewidth}
\centering
\includegraphics[width=1.0in]{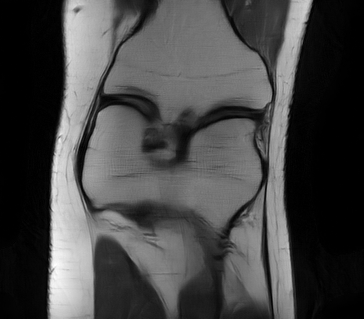}
\end{minipage}
}
\centering
\vspace*{-3mm}
\caption{\footnotesize{Example of MRI reconstruction using GAT, which leads to  the satisfactory reconstruction performance against the benign input $\mathbf z$ as well as the perturbed inputs $T(\mathbf z)$ and $\mathbf z + \boldsymbol\delta$.we rotate output at $T(\mathbf{z})$ $180^{\circ}$ for easier comparison with others.}} 
\label{fig: recon_visual_GAT}
\end{figure}

To solve \eqref{eq: nm_train_MMO_aug}, one major difficulty is how to select the  proper data transformation operations  $T(\cdot)$ to construct $\tilde{ \mathcal D}$. Motivated by the success of data augmentation in improving distributional robustness of image classification \citep{geirhos2018imagenet,hendrycks2019augmix}, we consider the following three representative data transformation operations: rotation \citep{taylor2018improving}, cutout \citep{devries2017improved}, and cutmix \citep{yun2019cutmix}; see Fig.\,\ref{fig: ill_exam} for an illustrative example. The rationale behind these augmentation operations is that invariant deep features used for image reconstruction can be learned against spatial perturbations,\textit{e.g.} , occlusion, translation, and rotation.
Our experiments in Sec.\,\ref{sec:experiments}  will show that the combination of these augmentation operations can lead to a superior robustness to input transformations. Meanwhile, the model learned from \eqref{eq: nm_train_MMO_aug}  (denoted by $\boldsymbol \theta_{\mathrm{GAT}}$)  also maintains robustness against adversarial input perturbations like AT (see Fig.\,\ref{fig: recon_visual_GAT}).

\vspace*{-5mm}
\section{Experiments}
\label{sec:experiments}

\vspace*{-1mm}
\subsection {Experiment Setup}
\paragraph*{Datasets and model architecture} In this paper, we focus on the 
Fully-sampled coronal proton density knee MRI dataset with 15-channel coils, which were obtained from the fastMRI database \citep{knoll2020fastmri}. 
The image size is  $320\times368$. In our paper, we used the unrolling-based image reconstruction model MoDL \citep{aggarwal2018modl}.
\vspace*{-3mm}
\paragraph*{Training setup} All of our models were trained in an end-to-end fashion using the Adam optimizer with  learning rate $1\times 10^{-4}$ and batch size $1$.
When implementing AT and GAT, we used the $10$-step PGD attack \citep{madry2017towards} with $\epsilon = 0.03/255$ for solving the inner maximization problem. When generating the transformed dataset $\tilde{\mathcal D}$ in GAT, we sampled one transformation operation for each original data sample from the random cutout, random cutmix, and rotate $180 ^{\circ}$ uniformly. And we choose the regularization parameter $\gamma = 3 $ in the implementation of GAT \eqref{eq: nm_train_MMO_aug}.  
\vspace{-3mm}
\paragraph*{Evaluation metrics}
In our paper, we used both normalized mean squared error (NMSE) and structural similarity index measure (SSIM) to evaluate the quality of reconstructed images. 
 The higher SSIM signifies  higher quality reconstruction results, while the higher NMSE represents the lower quality of image reconstruction.
To evaluate the robustness and accuracy of a model, we evaluate its performance when facing    three types of test-time inputs:
 \textbf{1) Benign test input} $\mathbf z$, which corresponds to the standard reconstruction accuracy;
\textbf{2) Adversarial test input} ($\mathbf z + \boldsymbol \delta$), which is generated   using the $20$-step PGD attack with $\epsilon = 0.03/255$ and corresponds to the adversarial robustness; 
 \textbf{3) Transformed test input $T(\mathbf z)$}: 
 model's robustness against input spatial transformations.

\subsection{Experiment results}
\begin{wraptable}{r}{76mm}
\vspace*{-2.4em}
\begin{center}
\caption{
\footnotesize{NMSE and SSIM results achieved by differently trained models, including NT (normal training), AT (adversarial training), and GAT (ours). The relative improvement/degradation with respect to NT is shown. }
}
\label{tablle:overall_performace}
\resizebox{0.48\textwidth}{!}{
\begin{tabular}{c|cc|cc|cc} 
\toprule[1pt]
\midrule
 \multicolumn{7}{c}{MRI reconstruction results on FastMRI}\\
\midrule 
%   \makecell{Perturbation \\ Radius} 
  \multirow{2}{*}{
    \begin{tabular}[c]{@{}c@{}}Method\end{tabular}
    } 
  &
  \multicolumn{2}{c|}{\begin{tabular}[c]{@{}c@{}}
  benign test input \\
  ($\mathbf z$)
\end{tabular}} &
\multicolumn{2}{c|}{
    \begin{tabular}[c]{@{}c@{}}
  adversarial test input \\
  ($\mathbf{z}+\boldsymbol \delta$) 
\end{tabular}} &
\multicolumn{2}{c}{
    \begin{tabular}[c]{@{}c@{}}
  transformed  test input \\
  ($T(\mathbf z)$) 
\end{tabular}} \\
%\hline
% \cline{2-7}
  & NMSE & SSIM & NMSE & SSIM & NMSE & SSIM  \\
  \midrule
  \text{NT}
    &\textbf{0.0013}  &\textbf{0.9442}  & 0.4952 & 0.4392 & 0.6824 & 0.7577 \\
  \text{AT}
  %\cite{liu2017voxelflow}
  &+0.0014 & -0.0478 &-\textbf{0.4900} & +\textbf{0.3827} & -0.5530 & +0.0354 \\
    \text{GAT} 
  %\cite{liu2017voxelflow}
  &+0.0017 & -0.0521&-0.4892 & +0.3748& -\textbf{0.6619}& +\textbf{0.0790} \\
\midrule
\bottomrule[1pt]
\end{tabular}
}
\end{center}
\vspace*{-1em}
\end{wraptable}

\paragraph*{Reconstruction accuracy and robustness}
We compared our proposed GAT with NT and AT. The overall quantitative results are shown in Tab.\,\ref{tablle:overall_performace}. As we can see, GAT is the most robust model against spatial transformation-based perturbations, without sacrificing much on accuracies against adversarial test inputs and benign  test inputs. 
These suggest that our proposed GAT improves robustness against adversarial perturbation and spatial transformation-based perturbation with a slight standard accuracy drop.

\begin{wraptable}{r}{.5\textwidth}
\vspace*{-2.4em}
\begin{center}

\caption{
\footnotesize{NMSE and SSIM results achieved by GAT under different data augmentation methods. The relative performance is shown with respect to NT. }
}
\label{tablle:diff_aug}

\resizebox{0.47\textwidth}{!}{
\begin{tabular}{c|cc|cc|cc} 
\toprule[1pt]
\midrule
 \multicolumn{7}{c}{Comparison of different augmentation methods}\\
\midrule 
  %\makecell{Perturbation \\ Radius} 
  \multirow{2}{*}{
    \begin{tabular}[c]{@{}c@{}}Method\end{tabular}
    } 
  &
  \multicolumn{2}{c|}{\begin{tabular}[c]{@{}c@{}}
  benign test input \\
  ($\mathbf z$)
\end{tabular}} &
\multicolumn{2}{c|}{
    \begin{tabular}[c]{@{}c@{}}
  adversarial test input \\
  ($\mathbf{z}+\boldsymbol \delta$) 
\end{tabular}} &
\multicolumn{2}{c}{
    \begin{tabular}[c]{@{}c@{}}
  transformed  test input \\
  ($T(\mathbf z)$) 
\end{tabular}} \\
%\hline
% \cline{2-7}
  & NMSE & SSIM & NMSE & SSIM & NMSE & SSIM  \\
  \midrule
  \text{NT}
    &\textbf{0.0013}  &\textbf{0.9442}  & 0.4952 & 0.4392 & 0.6824 & 0.7577 \\
  \text{GAT(cutout)}
  %\cite{liu2017voxelflow}
  & +0.0014 & -0.0502 &-\textbf{0.4896} & +\textbf{0.3824} &-0.6262 & +0.0777 \\
    \text{GAT(cutmix)} 
  %\cite{liu2017voxelflow}
  & +0.0014 &-0.0526 & -0.4894 & +0.3731 &-0.6183 & +0.0668 \\
      \text{GAT(rotation)} 
  %\cite{liu2017voxelflow}
  &+0.0028 & -0.0535&-0.4877 & +0.3180& -0.6593& +0.0476 \\
      \text{GAT} 
  %\cite{liu2017voxelflow}
  &+0.0017 & -0.0521&-0.4892 & +0.3748& -\textbf{0.6619}& +\textbf{0.0790} \\
\midrule
\bottomrule[1pt]
\end{tabular}}
\end{center}
\vspace*{-2em}
\end{wraptable}

\paragraph*{Effect of data augmentations} Next, we investigate how the choice of data augmentation affects model robustness.   Tab.\,\ref{tablle:diff_aug} shows 
the NMSE and SSIM performance of our approach (GAT) using different data augmentation strategies: cutout-only, cutmax-only,   rotation-only, and their combination. As we can see, each transformation operator used in GAT can improve the robustness against the spatial input transformations. However, the integration of all three transformation types into GAT achieves the best robustness at transformed test inputs. Meanwhile, we find that the use of cutout-only is able to achieve a great tradeoff with the improvement in adversarial robustness. 
\vspace*{-3mm}
\paragraph*{Visualization of reconstructed MRI images}  Fig.\,\ref{fig: recon_visual_2} shows examples of reconstructed MRI images using different approaches when facing different types of a test-time examples.
For ease of comparison, we apply  the inverse spatial transformation to the reconstruction result of a spatially-transformed input.  
As shown in Fig.\,\ref{fig: recon_visual_2}-(b), (d), and (e),   there exist less visible artifacts in the reconstruction results of our GAT-trained model. 
Compared to NT, both AT and GAT significantly improves model robustness against adversarial perturbations. While Fig.\,\ref{fig: recon_visual_2}-(a), (d), and (g) suggest that there might exist an accuracy-robustness tradeoff.
\section{Conclusion}
In this paper, we investigate the problem of adversarial robustness   of DL-based MRI image reconstruction.
We show that a deep image reconstruction model is highly susceptible to not only   pixel-level adversarial perturbations but also input spatial transformations, e.g., cropping and rotation. To improve the resilience of MRI image reconstruction, 
we  propose a generalized adversarial training method that is built upon a new robustness regularization metric by taking into account both adversarial perturbations and data transformations.
We show that our proposal is effective in a fastMRI dataset by comparing with  state-of-the-art baselines under a variety of evaluation metrics. In the future, we would like to generalize our approach to the self-supervised learning paradigm, and improve its computation efficiency through more lightweight robust training protocols.

{
\small
\bibliographystyle{unsrtnat}
\bibliography{refs}
}

\newpage

\appendix

\section{Appendix}
\setcounter{figure}{0}
\makeatletter 
\renewcommand{\thefigure}{A\arabic{figure}}% Figure counter representation
\renewcommand{\theHfigure}{A\arabic{figure}}% Hyperref figure hyperlink hook
\renewcommand{\thetable}{A\arabic{table}}
\renewcommand{\theHtable}{A\arabic{table}}

\makeatother
\setcounter{table}{0}

\setcounter{equation}{0}
\renewcommand{\theequation}{A\arabic{equation}}

\subsection{Transformation operations}
In this section, we provide some illustrations for different data transformation operations.
\begin{figure}[htbp]
\centering
\subfigure[\footnotesize{ cutout}]{
\begin{minipage}[t]{0.30\linewidth}
\centering
\includegraphics[width=1.0in]{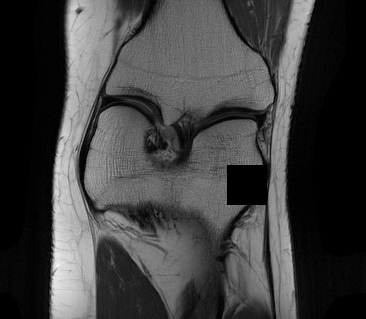}
\end{minipage}%
}%   
\subfigure[\footnotesize{cutmix}]{
\begin{minipage}[t]{0.30\linewidth}
\centering
\includegraphics[width=1.0in]{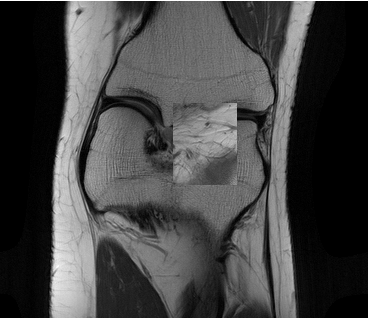}
\end{minipage}
}
\subfigure[\footnotesize{ rotation }]{
\begin{minipage}[t]{0.30\linewidth}
\centering
\includegraphics[width=1.0in]{Figures/transformed_ref.png}
\end{minipage}%
}%   
\vspace*{-3mm}
\caption{\footnotesize{Examples for different data transformation operations.
%using the AT model $f_{\boldsymbol \theta_{\mathrm{AT}}}$ at the spatially transformed input (by rotating $180 ^{\circ}$), the output image, and the reference image.
}} 
\label{fig: ill_exam}
\end{figure}
\subsection{Results visualization}

\begin{figure}[ht]
\centering

\subfigure[\footnotesize{ $f_{\boldsymbol \theta_{\mathrm{NT}}}(\mathbf z)$}]{
\begin{minipage}[t]{0.30\linewidth}
\centering
\includegraphics[width=1.0in]{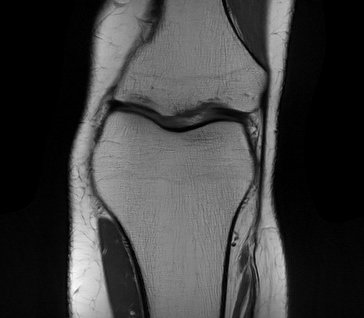}
\end{minipage}%
}%   
\subfigure[\footnotesize{$f_{\boldsymbol \theta_{\mathrm{NT}}}(T(\mathbf z))$ }]{
\begin{minipage}[t]{0.30\linewidth}
\centering
\includegraphics[width=1.0in]{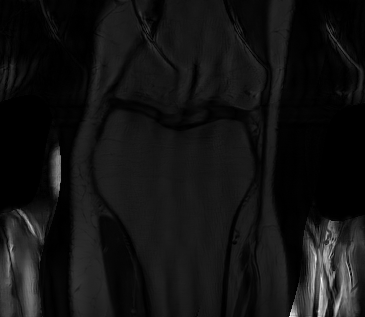}
\end{minipage}%
}%   
\subfigure[\footnotesize{$f_{\boldsymbol \theta_{\mathrm{NT}}}(\mathbf z +\boldsymbol \delta )$}]{
\begin{minipage}[t]{0.30\linewidth}
\centering
\includegraphics[width=1.0in]{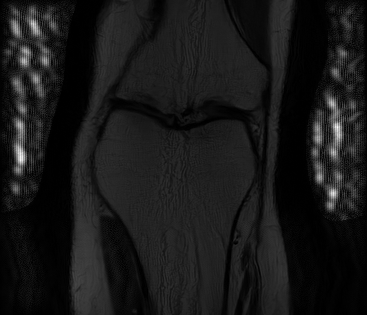}
\end{minipage}
}
\vspace*{-1mm}
\subfigure[\footnotesize{ $f_{\boldsymbol \theta_{\mathrm{AT}}}(\mathbf z)$}]{
\begin{minipage}[t]{0.30\linewidth}
\centering
\includegraphics[width=1.0in]{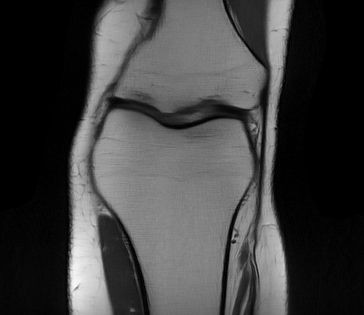}
\end{minipage}%
}%   
\subfigure[\footnotesize{$f_{\boldsymbol \theta_{\mathrm{AT}}}(T(\mathbf z))$ }]{
\begin{minipage}[t]{0.30\linewidth}
\centering
\includegraphics[width=1.0in]{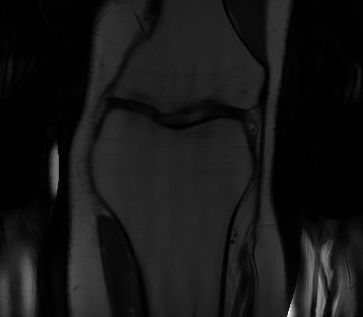}
\end{minipage}%
}%   
\subfigure[\footnotesize{$f_{\boldsymbol \theta_{\mathrm{AT}}}(\mathbf z +\boldsymbol \delta )$}]{
\begin{minipage}[t]{0.30\linewidth}
\centering
\includegraphics[width=1.0in]{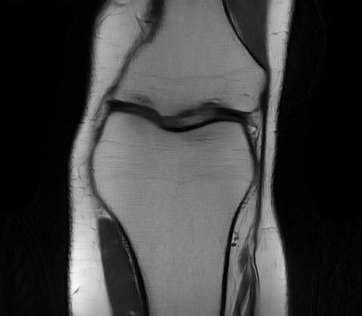}
\end{minipage}
}
\vspace*{-1mm}
\subfigure[\footnotesize{ $f_{\boldsymbol \theta_{\mathrm{GAT}}}(\mathbf z)$}]{
\begin{minipage}[t]{0.30\linewidth}
\centering
\includegraphics[width=1.0in]{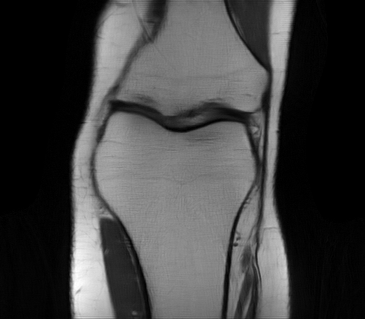}
\end{minipage}%
}%   
\subfigure[\footnotesize{$f_{\boldsymbol \theta_{\mathrm{GAT}}}(T(\mathbf z))$ }]{
\begin{minipage}[t]{0.30\linewidth}
\centering
\includegraphics[width=1.0in]{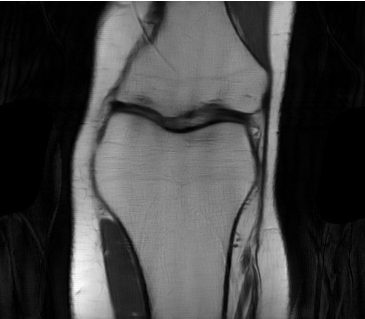}
\end{minipage}%
}%   
\subfigure[\footnotesize{$f_{\boldsymbol \theta_{\mathrm{GAT}}}(\mathbf z +\boldsymbol \delta )$}]{
\begin{minipage}[t]{0.30\linewidth}
\centering
\includegraphics[width=1.0in]{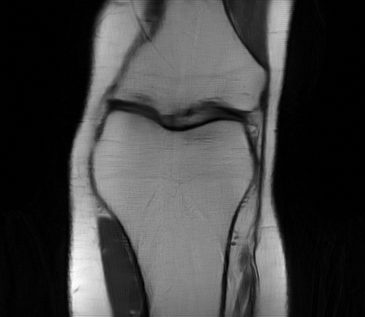}
\end{minipage}
}
\centering
\vspace*{-3mm}
\caption{\footnotesize{Visualization of MRI reconstructed images. Row 1: NT; Row 2:  AT; Row 3:    GAT.}}
\label{fig: recon_visual_2}
\end{figure}
\vspace{-4mm}

\end{document}

%% file: math_commands.tex
\usepackage{amsmath,amsfonts,bm}

\def\eqref#1{(\ref{#1})}
\def\1{\bm{1}}

\DeclareMathAlphabet{\mathsfit}{\encodingdefault}{\sfdefault}{m}{sl}
\SetMathAlphabet{\mathsfit}{bold}{\encodingdefault}{\sfdefault}{bx}{n}

\def\sC{{\mathbb{C}}}